\documentclass[12pt]{article}

\usepackage[T1]{fontenc}
\usepackage{libertine}
\usepackage[scaled=0.825]{beramono}

\usepackage{textcomp}
\usepackage{amsmath}
\usepackage{amssymb}

\usepackage{libertinust1math}

\usepackage{bm}
\usepackage{booktabs}
\usepackage{enumerate}
\usepackage{graphicx}
\usepackage[colorlinks=true]{hyperref}
\usepackage{lineno}
\usepackage[numbers,sort]{natbib}
\usepackage{siunitx}
\usepackage[dvipsnames]{xcolor}
\usepackage{xspace}

\makeatletter
\let\@fnsymbol\@arabic
\makeatother

\setcitestyle{citesep={;}}
\defcitealias{jcgm-2008s1}{JCGM~101:2008}

\hypersetup{linkcolor={red!50!black},
  citecolor={green!50!black},
  urlcolor={blue!80!black}}

\begin{document}

\title{Shades of Dark Uncertainty\\ and
  Consensus Value for the\\Newtonian Constant of Gravitation}
\author{Christos Merkatas\thanks{\texttt{christos.merkatas@nist.gov}}
\and 
Blaza Toman\thanks{\texttt{blaza.toman@nist.gov}}
\and
Antonio Possolo\thanks{\texttt{antonio.possolo@nist.gov}}
\and Stephan Schlamminger
\thanks{\texttt{stephan.schlamminger@nist.gov}\newline
National Institute of
  Standards and Technology, Gaithersburg, MD, USA}}
\date{May 23, 2019}
\maketitle

\newpage
\begin{abstract}
  The Newtonian constant of gravitation, $G$, stands out in the
  landscape of the most common fundamental constants owing to its
  surprisingly large relative uncertainty, which is attributable
  mostly to the dispersion of the values measured for it by different
  methods and in different experiments, each of which may have rather
  small relative uncertainty.

  This study focuses on a set of measurements of $G$ comprising
  results published very recently as well as older results, some of
  which have been corrected since the original publication. This set
  is inconsistent, in the sense that the dispersion of the measured
  values is significantly larger than what their reported
  uncertainties suggest that it should be. Furthermore, there is a
  loosely defined group of measured values that lie fairly close to a
  consensus value that may reasonably be derived from all the
  measurement results, and then there are one or more groups with
  measured values farther away from the consensus value, some
  appreciably higher, others lower.

  This same general pattern is often observed in many other
  interlaboratory studies and meta-analyses. In the conventional
  treatments of such data, the mutual inconsistency is addressed by
  inflating the reported uncertainties, either multiplicatively, or by
  the addition of ``random effects'', both reflecting the presence of
  \emph{dark uncertainty}. The former approach is often used by CODATA
  and by the Particle Data Group, and the latter is common in medical
  meta-analysis and in metrology. However, both achieve consistency
  ignoring how the measured values are arranged relative to the
  consensus value, and measured values close to the consensus value
  often tend to be penalized excessively, by such ``extra''
  uncertainty.

  We propose a new procedure for consensus building that models the
  results using latent clusters with different shades of dark
  uncertainty, which assigns a customized amount of dark uncertainty
  to each measured value, as a mixture of those shades, and does so
  taking into account both the placement of the measured values
  relative to the consensus value, and the reported uncertainties.  We
  demonstrate this procedure by deriving a new estimate for $G$, as a
  consensus value
  $G =
  \SI{6.67408e-11}{\meter\tothe{3}\kilo\gram\tothe{-1}\second\tothe{-2}}$,
  with
  $u(G) =
  \SI{0.00024e-11}{\meter\tothe{3}\kilo\gram\tothe{-1}\second\tothe{-2}}$.
\end{abstract}

\vspace{2pc}

\newpage
\section{Introduction}
\label{sec:introduction}

\hspace*{3em}{\fontsize{9}{11}\selectfont
    \begin{minipage}{0.575\linewidth}
      \begin{itemize}
      \item Gravitatem in corpora universa fieri, eamque
        proportionalem esse quantitati materi\ae{} in singulis
      \item Si Globorum duorum in se mutu\`{o} gravitantium materia
        undique, in regionibus qu\ae{} \`{a} centris \ae{}qualiter
        distant, homogenea sit: erit pondus Globi alterutrius in
        alterum reciproc\`{e} ut quadratum distanti\ae{} inter centra
      \end{itemize}

      \vspace*{-0.5ex}
      \hspace*{1.35em} I. Newton (1687) --- Matheseos Professore Lucasiano \\
      \hspace*{1.35em} \emph{Philosophi\ae{} Naturalis Principia
        Mathematica} \\ 
      \hspace*{1.35em} Liber Tertius: \emph{De Mundi Systemate}
    \end{minipage}}

\vspace*{1.75ex}

The NIST Reference on Constants, Units, and Uncertainty
(\url{https://physics.nist.gov/cuu/Constants/}) includes a list of 22
``Frequently used constants'', among them the Newtonian constant of
gravitation, $G$, which has the largest relative standard uncertainty
among these 22, by far, particularly after the values of the Planck
constant $h$, elementary electrical charge $e$, Boltzmann constant
$k$, and Avogadro constant $N_{\text{A}}$ were fixed in preparation
for the redefinition of the international system of units (SI)
\citep{newell-2018}. The constant $G$ appears as a proportionality
factor in Newton's law of universal gravitation and in the field
equations of Einstein's general theory of relativity
\citep{misner-2017}.

The surprisingly large uncertainty associated with $G$ is mostly an
expression of the dispersion of the values that have been measured for
it, which exceeds by far the reported uncertainties associated with
the individual measured values. \citet{rothleitner-2017} suggest that
``this gives reason to suspect hidden systematic errors in some of the
experiments. An alternative explanation is that although the values
are reported correctly, some of the reported uncertainties may be
lacking significant contributions. The uncertainty budgets can include
only what experimenters know and not what they do not know. This
missing uncertainty is sometimes referred to as a dark uncertainty''
\citep{thompson-2011}.

\citet{speake-2005} summarizes the role that $G$ plays in classical
and quantum physics, reviews the methods used to measure $G$ in
laboratory experiments, and discusses the outstanding challenges
facing such measurements, suggesting that improvements in the
measurement of length are key to reducing the uncertainty associated
with $G$, but also, somewhat discouragingly, suggesting that, owing to
``a multitude of subtle problems'', it may be a forlorn hope ever to
achieve mutual agreement to within 10 parts per million.

\citet{klein-2019} offers Modified Newtonian Dynamics (MOND)
\citep{milgrom-2015} as a striking and provocative explanation for why
some measured values should lie as far away from the currently
accepted consensus value \citep{mohr-2016} as they do, and shows how
they can be ``corrected.''

The principal aim of this contribution is to present a new approach to
derive a consensus value from the set of mutually inconsistent
measurement results for $G$ that the Task Group on Fundamental
Constants of the Committee on Data for Science and Technology (CODATA,
International Council for Science) used to produce its most recent
recommendation of a value for $G$ \citep{mohr-2016}, together with the
two, more recent measurement results reported by \citet{li-2018}. The
procedure we propose is equally applicable to similar reductions of
other, mutually inconsistent data sets obtained in interlaboratory
comparisons and in meta-analyses \citep{cooper-2009,baker-2013}.

In Section~\ref{sec:cavendish} we review a few, particularly
noteworthy measurements that directly or indirectly relate to $G$,
beginning with the measurement of the density of the Earth undertaken
by Henry Cavendish. Section~\ref{sec:consistency} is focused on the
evaluation of mutual consistency (or, homogeneity) of the measurement
results: we review several ways in which mutual consistency has
traditionally been gauged, and discuss how multiplicative and additive
statistical models may be used to produce consensus values when the
measurement results are mutually inconsistent.

Section~\ref{sec:shades} addresses a common complaint about the use of
models where dark uncertainty appears as a uniform penalty that
applies equally to all measurement results being combined into the
consensus value, regardless of whether the corresponding measured
values lie close or far from the consensus value, and motivates an
alternative approach.

This novel approach regards the measured values as drawings from a
mixture of probability distributions, effectively clustering the
measurements into subsets with different levels (\emph{shades}) of
dark uncertainty (Section~\ref{sec:mix}). If $n$ denotes the number of
measurements one wishes to blend, then we consider mixtures whose
number of components ranges from 1 to $n$, and use Bayesian model
selection criteria to identify the best
model. Section~\ref{sec:results} presents the results obtained by
application of the proposed model to the measurement results available
for $G$.

The conclusions, presented in Section~\ref{sec:conclusions}, include
the observation that advances in the measurement of $G$ involve not
only substantive developments in measurement methods, but also in the
statistical modeling that informs productive data reductions and
enables realistic uncertainty evaluations.

\section{Historical retrospective}
\label{sec:cavendish}

\citet[Page~520]{cavendish-1798} lists 29 determinations of the
relative density (or, specific gravity) $d_{\oplus}$ of the Earth. The
first 6, produced in the experiments of August 5-7, 1797, were made
using one particular wire to suspend the wooden arm of the apparatus
bearing two small leaden balls. Cavendish found that this wire was
insufficiently stiff, and he replaced it with a stiffer wire for the
23 determinations between August 12, 1797 and May 30, 1798
\citep[Page~485]{cavendish-1798}.

This second group of 23 determinations has average
\SI{5.480}{\gram/\centi\meter\cubed}. Cavendish points out that the
range of these determinations is \SI{0.75}{\gram/\centi\meter\cubed},
``so that the extreme results do not differ from the mean by more than
0.38, or $\frac{1}{14}$ of the whole, and therefore the density should
seem to be determined hereby, to great exactness''
\citep[Page~521]{cavendish-1798}. Following a brief recapitulation of
sources of error discussed amply earlier in the paper,
\citet[Page~522]{cavendish-1798} concludes that ``it seems very
unlikely that the density of the earth should differ from 5.48 by so
much as $\frac{1}{14}$ of the whole.''

Since the standard deviation of those 23 determinations is
\SI{0.19}{\gram/\centi\meter\cubed}, the aforementioned
\SI{0.38}{\gram/\centi\meter\cubed} (``$\frac{1}{14}$ of the whole'')
may be regarded as an expanded uncertainty for approximate 95\,\%
coverage. (Also in agreement with the conventional, crude estimate of
the standard deviation as one fourth of the range, that is
$(\SI{0.75}{\gram/\centi\meter\cubed})/4$ $\approx$
\SI{0.19}{\gram/\centi\meter\cubed} in this case \citep{hozo-2005}.)

In other words, Cavendish seems effectively to have regarded the 23
determinations made using the second wire as a sample from the
distribution of the measurand, and used an assessment of their
dispersion as evaluation of what nowadays we would call standard
uncertainty, rather than using anything like the standard deviation of
the average of the same 23 determinations, which would have been
$\sqrt{23} \approx 4.8$ times smaller than
\SI{0.19}{\gram/\centi\meter\cubed}. (It should be noted that none of
the terms \emph{probable error} \citep{bessel-1815}, \emph{mean error}
\citep{gauss-1823-bd4}, \emph{standard deviation}
\citep{pearson-1894}, or \emph{standard error} \citep{yule-1897} were
in use at the time.)

The mass $m$ of an object lying on the surface of the ellipsoid
defined in the World Geodetic System (WGS~84) \citep{nima-2000}, at
geodetic latitude $\varphi$, satisfies
$m g(\varphi) = G M_{\oplus} m / r^{2}(\varphi)$, where $G$ is the
Newtonian constant of gravitation, $M_{\oplus}$ denotes the mass of
the Earth, $g(\varphi)$ denotes the theoretical acceleration due to
gravity (exclusive of the effect of the centrifugal acceleration due
to the Earth's rotation), and $r(\varphi)$ denotes the Earth's
geocentric radius at latitude $\varphi$. If $R_{3} =
\SI{6371000.79}{\meter}$ denotes the radius of a sphere with the same
volume as the WGS~84 ellipsoid \citep{moritz-2000}, then $M_{\oplus} =
(4/3) \pi R_{3}^{3} d_{\oplus}$. Therefore,
$G = 3 g(\varphi) r^{2}(\varphi) / (4 \pi R_{3}^{3} d_{\oplus})$.

Substituting $d_{\oplus} = \SI{5480}{\kilo\gram/\meter\cubed}$ as
measured by Henry Cavendish, and
$g(\varphi) = \SI{9.812004}{\meter/\second\squared}$ and
$r(\varphi) = \SI{6365097}{\meter}$ for the latitude,
$\varphi = \SI{51.4578}{\degree}\,\text{N}$, of Clapham Common, South
London, where his laboratory was located, and neglecting the elevation
above sea level of the same location (approximately \SI{30}{\meter}),
yields $G_{\text{C}} = \SI{6.69693e-11}%
{\meter\cubed\kilo\gram\tothe{-1}\second\tothe{-2}}$. (Note that the
subscript ``C'' that is used here serves only to indicate the
provenance of this estimate of $G$, not to suggest that the true value
of the constant depends on location.) The foregoing value for
$g(\varphi)$ was computed according to
\citet[Equation~(4-1)]{nima-2000}, and the radius $r(\varphi)$ was
computed using the lengths of the semi-major and semi-minor axis of
the WGS~84 ellipsoid listed in \citet[Tables~3-1, 3-3]{nima-2000}.

Since $u(d_{\oplus})/d_{\oplus} = 3.5\,\%$ and we take the geometry of
WGS 84, and the latitude of Clapham Common, as known quantities, this
is also the relative uncertainty associated with $G_{\text{C}}$.
More impressive still is the fact that the error in $G_{\text{C}}$,
relative to the CODATA 2014 recommended value,
$G_{2014} = \SI{6.67408e-11}%
{\meter\cubed\kilo\gram\tothe{-1}\second\tothe{-2}}$
\citep{mohr-2016}, is only 0.34\,\%. The comparable relative
``errors'' associated with the contemporary measured values listed in
Table~\ref{tab:newtonian} range from $-0.033\,\%$ to $0.023\,\%$,
indicating that, in the intervening 220 years, the worst relative
``error'' in the determination of $G$ has been reduced by no more than
10-fold.

$G$ was of no concern to Cavendish, and neither did Newton introduce
it in the \emph{Principia} \citep{newton-1687}. More than 70 years
would have to elapse after Cavendish ``weighed the Earth'', before
even a particular symbol would be advanced for the gravitational
constant --- and the symbol at first was ``$f$'', not ``$G$''
\citep{cornu-1873}.

According to Hartmut Petzold (formerly with the Deutsches Museum,
Munich, \emph{personal communication}), the birthday of the expression
``gravitational constant'' was on one of these three days, December
16-18, 1884: on December 16th, Arthur K\"{o}nig and Franz Richarz
submitted a handwritten proposal to measure ``the mean density of the
earth''; two days later Helmholtz presented their proposal to the
Royal Prussian Academy of Sciences in Berlin with the modified title
``A new method for determining the gravitational constant''
\citep{konig-1884}.

In the evening session of June 8, 1894, of the Royal Institution of
Great Britain, Charles Vernon Boys also used the symbol $G$ when he
made a presentation on the Newtonian constant of gravitation, and
announced \SI{6.6576e-11}%
{\meter\tothe{3}\kilo\gram\tothe{-1}\second\tothe{-2}} as ``adopted
result'' derived from experiments using gold and lead balls in a
torsion balance \citep{boys-1894}. The relative difference between
this determination and CODATA's $G_{2014}$ is $-0.25\,\%$.

\begin{table}
  \centering
  \begin{tabular}{lSSl} \midrule
    & {$G$} & {$u(G)$} \\
    & \multicolumn{2}{c}{\small%
      $/ 10^{-11} \si{\meter\tothe{3}\kilo\gram\tothe{-1}%
      \second\tothe{-2}}$} \\
    \cmidrule{2-3}
    NIST-82     & 6.67248  & 0.00043  & \citep{luther-1982}\\          
    TR\&D-96    & 6.6729   & 0.00050  & \citep{karagioz-1996}\\        
    LANL-97     & 6.67398  & 0.00070  & \citep{bagley-1997} \\         
    UWash-00    & 6.674255 & 0.000092 & \citep{gundlach-2000} \\       
    BIPM-01     & 6.67559  & 0.00027  & \citep{quinn-2001} \\          
    UWup-02     & 6.67422  & 0.00098  & \citep{kleinevoss-2002} \\     
    MSL-03      & 6.67387  & 0.00027  & \citep{armstrong-2003} \\      
    HUST-05     & 6.67222  & 0.00087  & \citep{luo-1998,hu-2005} \\     
    UZur-06     & 6.67425  & 0.00012  & \citep{schlamminger-2006} \\   
    HUST-09     & 6.67349  & 0.00018  & \citep{tu-2010} \\             
    JILA-10     & 6.67260  & 0.00025  & \citep{parks-2010} \\          
    BIPM-14     & 6.67554  & 0.00016  & \citep{quinn-2013,quinn-2014} \\
    LENS-14     & 6.67191  & 0.00099  & \citep{rosi-2014} \\           
    UCI-14      & 6.67435  & 0.00013  & \citep{newman-2014} \\         
    HUST-TOS-18 & 6.674184 & 0.000078 & \citep{li-2018} \\
    HUST-AAF-18 & 6.674484 & 0.000078 & \citep{li-2018} \\
  \end{tabular}
  \caption{Measurement results for $G$ used in this study. The top
    fourteen lines reproduce the entries in
    \citet[Table~XV]{mohr-2016}, except for JILA-10, which has
    meanwhile been corrected as described in the text. The bottom two
    lines contain the results reported by \citet{li-2018}, obtained
    using the time-of-swing (TOS) method and the
    angular-acceleration-feedback (AAF) method for the torsion
    pendulum, which have the smallest associated uncertainties
    achieved thus far
    \citep{schlamminger-2018}.} \label{tab:newtonian}
\end{table}

In this study we focus on the set of measurement results listed in
Table~\ref{tab:newtonian}, which includes the results that CODATA used
to produce the 2014 recommended value for $G$, together with two, more
recent determinations. Since some of these results differ from their
originally published versions, the following remarks clarify the
precise provenance of all the measurement results listed. For the sake
of brevity, we use the scale factor
$\gamma = 10^{-11} \si{\meter\tothe{3}%
  \kilo\gram\tothe{-1}\second\tothe{-2}}$.

\begin{description}
\item[NIST-82] The result published originally,
  $G/\gamma=6.6726\pm 0.0005$ \citep{luther-1982}, had not been
  corrected for an effect caused by the anelasticity of the torsion
  fiber. The corresponding result listed in Table~\ref{tab:newtonian}
  reflects an anelasticity correction applied by CODATA. It should be
  noted that the change in the reported uncertainty (down from 0.0005
  in the original publication, to the 0.00043 in
  Table~\ref{tab:newtonian}) is not a consequence of this correction
  but results from a refinement of the uncertainty analysis that the
  authors did between the time when the result was first published and
  when it was used for the 1986 adjustment of the fundamental physical
  constants \citep{cohen-1987}.
\item[TR\&D-96] Identical to the published measurement result
  \citep{karagioz-1996}.
\item[LANL-97] In 2010, CODATA corrected the result published
  originally, $G/\gamma = 6.6740\pm 0.0007$ \citep{bagley-1997} to take
  into account uncertainties in the measurement of the quality factor
  of the torsion pendulum. The quality factor is needed to calculate
  the correction caused by the anelastic properties of the fiber.
\item[UWash-00] The measured value listed in the original
  work~\citep{gundlach-2000}, $G/\gamma = 6.674\,215\pm 0.000\,092$, is
  $6\times10^{-6}$ lower than the value used by CODATA. After the
  result was published, the authors noticed the omission of a small
  effect and communicated a corrected value to CODATA. The small
  effect was caused by a a mass that is mounted on the top of the
  torsion fiber, and is itself suspended by a thicker fiber. In this
  experiment, the gravitational torque is counteracted by the inertia
  of the pendulum in an accelerated rotating frame. The acceleration
  acts also on the pre-hanger, and its effect must be taken into
  account.  No erratum is publicly available.
\item[BIPM-01] Identical to the published result~\citep{quinn-2001}.
\item[UWup-02] Identical to the published result \citep{kleinevoss-2002}.
\item[MSL-03]  Identical to the published result~\citep{armstrong-2003}.
\item[HUST-05] The measurement result published originally in 1999,
  $G/\gamma = \num{6.6699} \pm \num{0.0007}$ \citep{luo-1998}, differs
  appreciably from the corresponding result used by CODATA. The
  measured value is lower than its CODATA counterpart, with a relative
  difference of $3.5\times 10^{-4}$. However, two needed corrections
  had not been applied: first, for the gravitational effect of the air
  that is displaced by the field masses; second, for the density
  inhomogeneity of the source masses. The result, as updated in 2005,
  became $G/\gamma = 6.672\,3\pm 0.000\,9$~\citep{hu-2005}, where the
  updated measured value is larger than CODATA's, the relative
  difference being $1.1\times 10^{-5}$.  In 2014, CODATA applied a
  third correction for the anelasticity of the fiber.
\item[UZur-06] Identical to the published result~\citep{schlamminger-2006}.
\item[HUST-09] Identical to the published result~\citep{tu-2010}.
\item[JILA-10] The authors of the original work~\citep{parks-2010},
  which listed $G/\gamma = \num{6.67234} \pm \num{0.00014}$ as
  measurement result, realized that two effects had been
  miscalculated. In 2018, they sent an erratum to CODATA reporting a
  corrected value of $G/\gamma = \num{6.67260} \pm
  \num{0.00025}$. First, the pendulum bob rotates under excursion from
  the equilibrium position due to a differential stretching of the
  support wire. The rotation is different in the calibration mode from
  the measurement mode. The second effect also has to do with the
  rotation of the bob. If the laser beam is not perfectly centered on
  the mass centers, a rotation can cause an apparent length change
  (Abbe effect). The Abbe effect was not properly calculated in the
  initial publication. These two effects have different signs,
  yielding a final result that differs relatively only by
  $+3.9\times 10^{-5}$ from the value in the original publication. An
  erratum has been submitted for publication in Physical Review
  Letters.
\item[BIPM-14] The measurement result reported originally
  $G/\gamma = \num{6.67545} \pm \num{0.00018}$ \citep{quinn-2013} was
  superseded by the result listed in an erratum published in
  2014~\citep{quinn-2014}. This was the value used by CODATA. The
  relative change in value of $-13.5\times 10^{-5}$ was caused by the
  density inhomogeneity of the source masses. In the original
  publication, the corresponding correction had inadvertently been
  applied twice.
\item[LENS-14] Identical to the published result \citep{rosi-2014}.
\item[UCI-14] In the original publication~\citep{newman-2014}, the
  authors reported a slightly (by $3\times 10^{-6}$) smaller value,
  $G/\gamma = 6.674\,33\pm 0.000\,13$. The reported value is an
  average of three measurements. The authors used an unweighted
  average, while CODATA used a weighted average and considered the
  correlation between the three results.
\item[HUST-TOS-18] Identical to the published result \citep{li-2018}.
\item[HUST-AAF-18] Identical to the published result \citep{li-2018}.
\end{description}

\section{Mutual consistency}
\label{sec:consistency}

A set of measurement results, comprising pairs of measured values and
associated standard uncertainties, for example $\{(G_{j}, u(G_{j}))\}$
as in Table~\ref{tab:newtonian}, is said to be mutually consistent
(or, \emph{homogeneous}) when the variability of the measured values
is statistically comparable to the reported uncertainties: for
example, when the standard deviation of the $\{G_{j}\}$ is practically
indistinguishable from the ``typical'' $\{u(G_{j})\}$ (say, their
median).

The standard deviation of the $\{G_{j}\}$ in Table~\ref{tab:newtonian}
is \SI{0.00109}{\meter\tothe{3}\kilo\gram\tothe{-1}\second\tothe{-2}},
while the median of the $\{u(G_{j})\}$ is
\SI{0.00026}{\meter\tothe{3}\kilo\gram\tothe{-1}\second\tothe{-2}}:
the former is 4.2 times larger than the latter, indicating that
the measured values are much more dispersed than their associated
uncertainties suggest they should be.

This implies that either the different experiments are measuring
different measurands, or there are sources of uncertainty yet
unrecognized that are not expressed in the reported uncertainties.  If
the different experiments indeed are measuring the same measurand,
then these uncertainties are much too small, and the lurking, yet
unrecognized ``extra'' component is what \citet{thompson-2011}
felicitously have dubbed \emph{dark uncertainty} because it is
perceived only once independent results are inter-compared. Dark
uncertainty may derive from a single or from multiple sources of
uncertainty.

Cochran's $Q$ test, which is the conventional chi-squared test of
mutual consistency, is very widely used, even if it suffers from
important limitations and misunderstandings \citep{hoaglin-2016}.  For
the measurement results in Table~\ref{tab:newtonian}, the test
statistic is $Q=198$ on 15 degrees of freedom: since the reference
distribution is chi-squared with 15 degrees of freedom,
$\chi_{15}^{2}$, the $p$-value of the test is essentially zero, hence
the conclusion of heterogeneity.

\subsection{Multiplicative models}
\label{sec:multiplicative}

\citet{birge-1932} suggested an approach for the combination of
mutually inconsistent measurement results that involves: first,
inflating the reported standard uncertainties using a multiplicative
inflation factor $\kappa$ sufficiently large to make the results
mutually consistent; second, combining the measured values into a
weighted average whose weights are inversely proportional to the
squared uncertainties. (Note that the value of the inflation factor
does not affect the value of the estimate of $G$, only its associated
uncertainty.)

The inflation factor is commonly set equal to the \emph{Birge Ratio},
$\kappa = R_{\text{B}}= \big[\sum_{j=1}^{n} w_{j} (G_{j} -
\overline{G})^{2} / (n-1)\big]^{\text{\textonehalf}} = 3.6$, where
$n=16$ denotes the number of measurement results and $\overline{G}$
denotes their weighted average corresponding to weights
$\{w_{j}=1/u^{2}(G_{j})\}$. This choice of value for $\kappa$ makes
Cochran's statistic equal to its expected value, hence is a
method-of-moments estimate. Birge's approach is used routinely by the
Particle Data Group (\url{pdg.lbl.gov}) \citep{tanabashi-2018}, and
also by CODATA to produce recommended values for some of the
fundamental physical constants \citep{mohr-2016}, $G$ in particular.

The inflation factor $\kappa$ may be determined in many other
ways. For example, as the smallest multiplier for the $\{u(G)\}$ that
yields a value of the chi-squared statistic as large as possible yet
shy of the critical value for the test. For the data in
Table~\ref{tab:newtonian}, and for a test whose probability of Type I
error is 0.05 (the probability of incorrectly rejecting the hypothesis
of homogeneity), the critical value is 24.996, and the corresponding,
smallest inflation factor that achieves homogeneity is
$\kappa = 2.813$.

The statistical model underlying the multiplicative adjustment of the
uncertainties regards the $j$th measured value as the true value $G$
plus an error commensurate with $u(G_{j})$ and magnified by
$\kappa$. More precisely, as $G_{j} = G + \kappa \varepsilon_{j}$,
where the $\{\varepsilon_{j}\}$ are modeled as non-observable outcomes
of independent Gaussian random variables all with mean 0 but with
standard deviations $\{u(G_{j})\}$. The consequence is that the
effective measurement errors $\{\kappa \varepsilon_{j}\}$ are then
Gaussian random variables with standard deviations
$\{\kappa u(G_{j})\}$.

The two choices of $\kappa$ reviewed above appear reasonable but are
\emph{ad hoc} (yet another \emph{ad hoc} choice is discussed in
Section~\ref{sec:residuals}). A principled, and generally preferable
alternative is maximum likelihood estimation, whereby the ``optimal''
consensus value $G$ and inflation factor $\kappa$ maximize a product
of Gaussian densities evaluated at the measured values $\{G_{j}\}$,
all with the same mean $G$, and standard deviations
$\{\kappa u(G_{j})\}$. The idea here is to select values for $G$ and
$\kappa$ that render the data ``most likely.''

\begin{figure}
  \centering
  \includegraphics[keepaspectratio=true,width=\linewidth]%
  {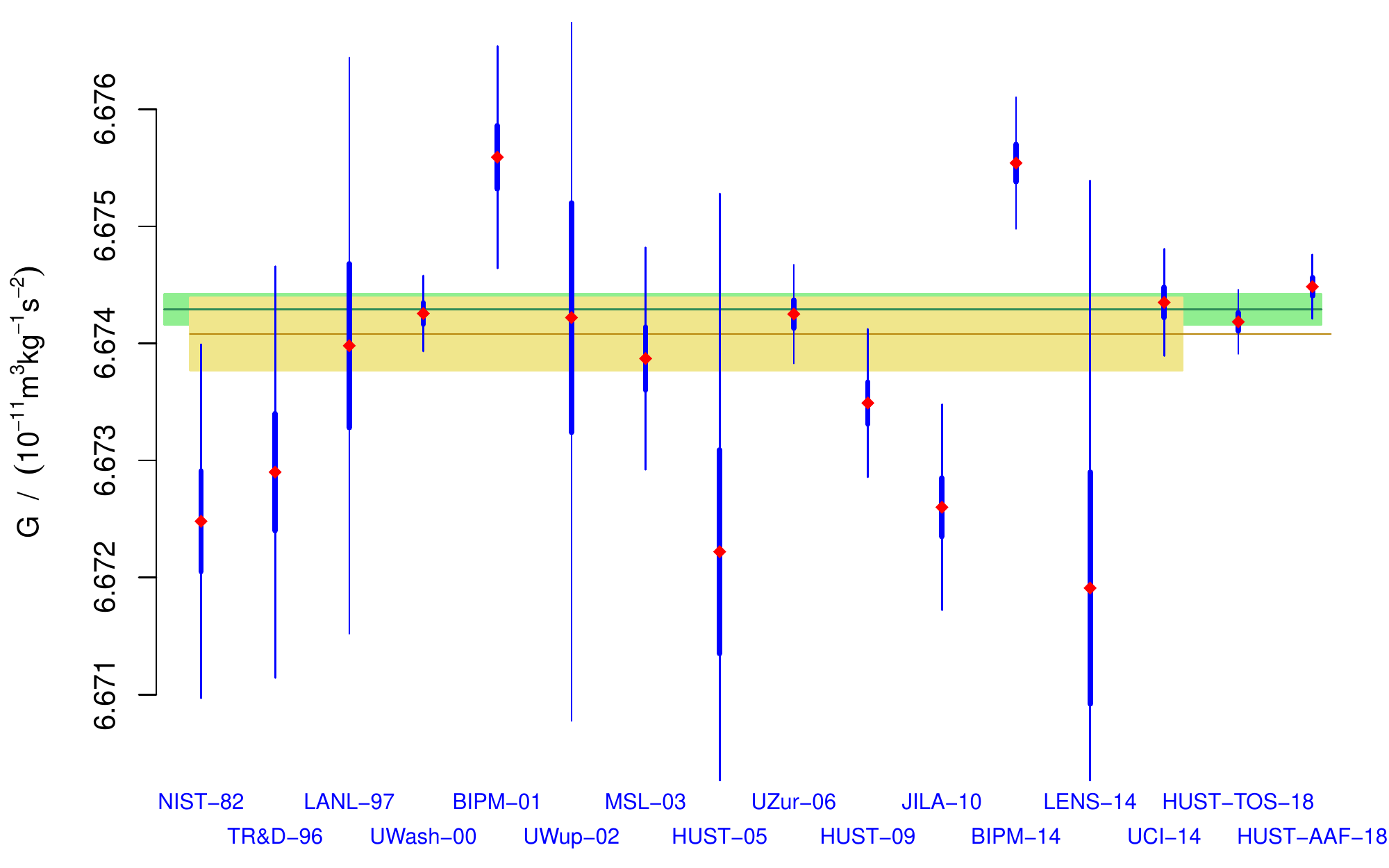}
  \caption{Measurement results from Table~\ref{tab:newtonian}, where
    the measured values are represented by red diamonds. The thick
    vertical blue line segments represent $\{G_{j} \pm
    u(G_{j})\}$. The thin line segments, several of which are
    truncated, represent $\{G_{j} \pm \widehat{\kappa} u(G_{j})\}$,
    where $\widehat{\kappa} = 3.5$ is the maximum likelihood estimate
    of the inflation factor. The horizontal green line represents the
    consensus value $\widehat{G}$, and the light green band represents
    $\widehat{G} \pm u(\widehat{G})$. The horizontal brown line and
    light brown band are the counterparts of the green line and light
    green band, for the CODATA recommended value $G_{2014}$, which did
    not incorporate the measurement results from either
    \textsf{HUST-TOS-18} or \textsf{HUST-AAF-18}, and used the
    uncorrected result from \textsf{JILA-10}. Compare with
    Figure~\ref{fig:dl}.} \label{fig:birge}
\end{figure}

The maximum likelihood estimates derived from the data in
Table~\ref{tab:newtonian}, are $\widehat{G} = \SI{6.67429e-11}%
{\meter\tothe{3}\kilo\gram\tothe{-1}\second\tothe{-2}}$ and
$\widehat{\kappa} = 3.5$. The evaluations of the associated
uncertainties, obtained using the parametric statistical bootstrap
\citep{efron-1993}, are $u(\widehat{G}) =$ \SI{0.00013e-11}%
{\meter\tothe{3}\kilo\gram\tothe{-1}\second\tothe{-2}}
(Table~\ref{tab:comparison}, row BRM), and
$u(\widehat{\kappa}) = 0.6$. A 95\,\% coverage interval for $\kappa$
ranges from 2.2 to 4.6, thus suggesting that any estimate of the
inflation factor $\kappa$ is bound to be clouded by very substantial
uncertainty.  Figure~\ref{fig:birge} depicts the results.

\citet{rothleitner-2017} sound a note of despair at the conclusion of
their review of the history, status, and prospects for improvement of
the measurements of $G$: ``Given the current situation in the
measurement of $G$, it is difficult to see how our knowledge of $G$
can be improved, for example, $\chi^{2}$ will not decrease by adding
new experiments, as it is a sum of squares and can increase only with
new data. The Birge ratio can decrease by increasing $\sqrt{N-1}$ in
the denominator; however, this will be a slow process. If an
additional 13 experiments are performed (which could take another 30
years if past experiments are an indication), $R_{\text{B}}$ can be
reduced by a factor 1.4 if the values are close to the current average
value. It is equally difficult to see how the multiplicative factor
that CODATA used to bring all normalized residuals below two can be
decreased. Thus, decreasing the current uncertainty assigned to the
recommended value of G does not seem to be possible --- at least, not
in the foreseeable future.''

Although we agree that reducing the uncertainty associated with $G$ is
an outstanding challenge in precision measurement, we believe that
conventional metrics for mutual inconsistency, be they Cochran's $Q$
or the Birge ratio, are not the most informative means to gauge
progress or lack thereof, and that more productive avenues for data
reduction are available as we shall illustrate forthwith. Furthermore,
in Section~\ref{sec:residuals} we show that the goal of bringing ``all
normalized residuals below two'' is excessively restrictive, hence
ought not to be used as a quality criterion whereon to judge the
mutual consistency of any collection of measurement results.

\subsection{Normalized residuals}
\label{sec:residuals}

\citet{mohr-2000} introduce the notion of ``normalized residual'' in
the context of the nonlinear least squares method that CODATA has been
using to derive adjusted values of the fundamental constants
$z_{1}, \dots, z_{M}$ from a collection of measured values
$q_{1}, \dots, q_{N}$ of quantities that are functionally related to
those constants by measurement equations
$\{q_{i} = f_{i}(z_{1}, \dots, z_{M})\}$, where the $\{f_{i}\}$ are
determined by the laws of physics and $N > M$. The normalized residual
corresponding to $q_{i}$ is $r_{i} = (q_{i} - \widehat{q}_{i})/u_{i}$,
with $u_{i}=u(q_{i})$ the standard uncertainty associated with
$q_{i}$, and
$\widehat{q}_{i} = f_{i} (\widehat{z}_{1}, \dots, \widehat{z}_{M})$,
where $\widehat{z}_{1}, \dots, \widehat{z}_{M}$ denote the adjusted
values of the constants.

For the 2014 adjustment of the value of $G$, ``the Task Group decided
that it would be more appropriate to follow its usual approach of
treating inconsistent data, namely, to choose an expansion factor that
reduces each $|r_{i}|$ to less than 2'' \citep{mohr-2016}. The idea is
aligned with Birge's approach, that $u_{i}$ should be replaced by
$\kappa u_{i}$, where $\kappa$ is the aforementioned expansion factor,
thereby reducing the magnitude of the residuals.
The adjusted value is the weighted average of the $\{G_{j}\}$, with
weights proportional to $1/(\kappa u_{i})^{2}$.

Applying this same procedure to the measurement results listed in
Table~\ref{tab:newtonian} yields an expansion factor $\kappa =
3.9$. The corresponding 
estimate of $G$ is
\SI{6.67429e-11}{\meter\tothe{3}\kilo\gram\tothe{-1}\second\tothe{-2}},
with associated standard uncertainty \num{0.00015e-11}
\si{\meter\tothe{3}\kilo\gram\tothe{-1}\second\tothe{-2}}
(Table~\ref{tab:comparison}, row MTE).

\begin{table}
  \centering
  {\fontsize{8}{10}\selectfont
  \begin{tabular}{lSclScclSclS} \cmidrule{1-5} \cmidrule{8-12}
    & {$r$} &&       & {$r$} &&&   & {$r^{\ast}$} &&    & {$r^{\ast}$}\\
    \cmidrule{1-2} \cmidrule{4-5} \cmidrule{8-9} \cmidrule{11-12}
 NIST-82&-4.2  &&UZur-06    &-0.32&&& NIST-82&-4.2  &&    UZur-06&-0.34\\
TR\&D-96&-2.8  &&HUST-09    &-4.4 &&&TR\&D-96&-2.8  &&    HUST-09&-4.5\\
 LANL-97&-0.44 &&JILA-10    &-6.8 &&& LANL-97&-0.44 &&    JILA-10&-6.8\\
UWash-00&-0.37 &&BIPM-14    & 7.8 &&&UWash-00&-0.40 &&    BIPM-14& 8.0\\
 BIPM-01& 4.8  &&LENS-14    &-2.4 &&& BIPM-01& 4.9  &&    LENS-14&-2.4\\
 UWup-02&-0.070&&UCI-14     & 0.47&&& UWup-02&-0.070&&     UCI-14& 0.49\\
  MSL-03&-1.6  &&HUST-TOS-18&-1.3 &&&  MSL-03&-1.6  &&HUST-TOS-18&-1.5\\
 HUST-05&-2.4  &&HUST-AAF-18& 2.5 &&& HUST-05&-2.4  &&HUST-AAF-18& 2.9\\
    \cmidrule{1-5} \cmidrule{8-12}
\end{tabular}}
\caption{Normalized residuals computed according to the conventional
  definition (left panel), and involving the correct denominator
  (right panel).} \label{tab:residuals}
\end{table}

\begin{figure}
  \centering
  \includegraphics[keepaspectratio=true,width=0.85\linewidth]%
  {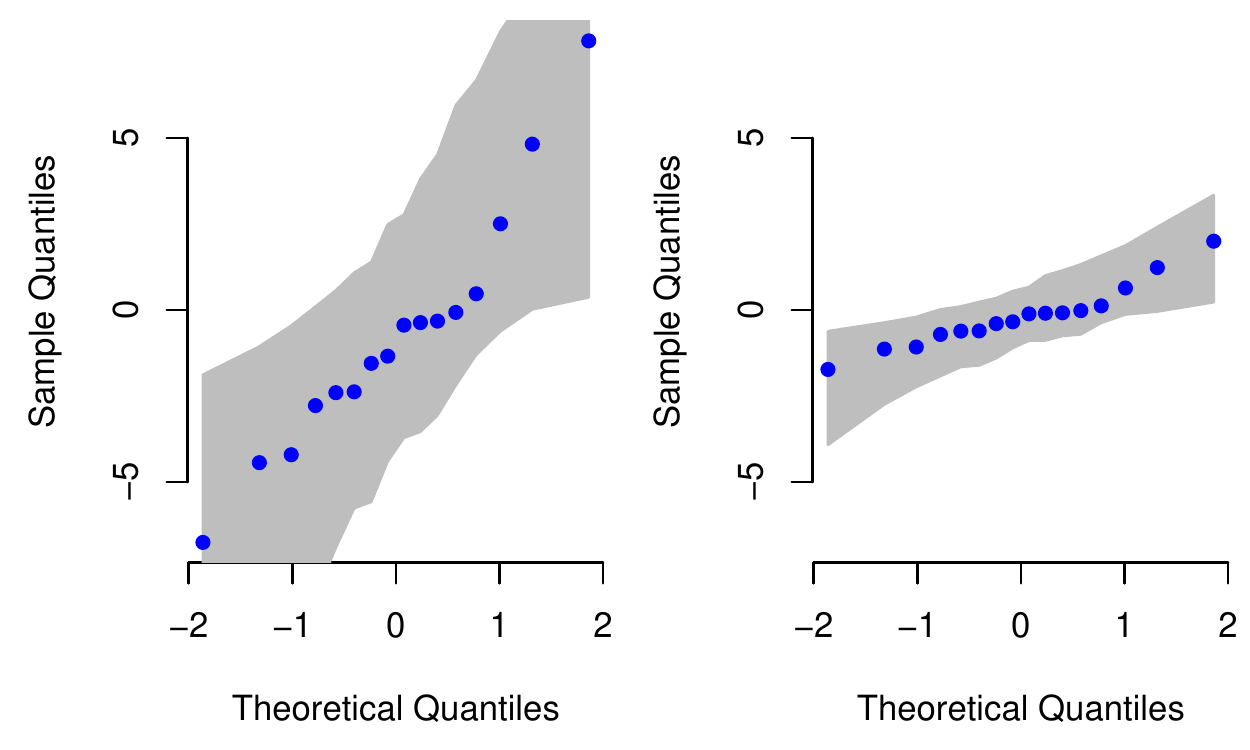}
  \caption{QQ-plots \citep{wilk-1968,chambers-1983} of the normalized
    residuals before (left panel) and after (right panel) expansion of
    the $\{u(G_{j})\}$ as described in
    Section~\ref{sec:residuals}. The abscissas of the points are
    approximately equal to the values expected for the smallest,
    second smallest, etc. in a Gaussian sample of this size. The
    ordinates are the smallest, second smallest, etc., of the
    residuals. If the models fit the data perfectly, then the dots in
    each plot should all fall on a straight line: the gray bands
    account for sampling variability, and the models are deemed to be
    adequate for the data when the dots all lie inside the gray
    bands. The vertical scale is the same for the two plots, showing
    that the expansion of the $\{u(G_{j})\}$ reduces the sizes of the
    residuals markedly.} \label{fig:residuals}
\end{figure}

This approach is justified by the belief that the $\{r_{i}\}$ should
be approximately like a sample from a Gaussian distribution, which
Figure~\ref{fig:residuals} indeed supports. There are, however, two
issues with this approach to achieve mutual consistency.

First, and this is the minor issue, the denominator of
$r_{j} = (G_{j} - \overline{G})/u_{j}$ should be
$u(G_{j}-\overline{G})$, not $u(G_{j})$, because the latter does not
recognize the uncertainty associated with $\overline{G}$ or the
correlation between $G_{j}$ and
$\overline{G}$. Table~\ref{tab:residuals} lists the values of the
normalized residuals $\{r_{j}\}$ as defined conventionally, and their
counterparts
$\{r_{j}^{\ast} = (G_{j} - \overline{G})/u(G_{j}-\overline{G})\}$
involving the correct denominator (evaluated using the parametric
statistical bootstrap \citep{efron-1993}). The differences between
corresponding values indeed are minor and largely inconsequential in
this case.

Second, and this is the major issue, if the (properly) normalized
residuals indeed are like a sample of size $n$ from a Gaussian
distribution with mean 0 and standard deviation 1, then, according to
the Fisher-Tippett-Gnedenko theorem \citep{gumbel-2004}, the expected
value of the largest residual is approximately equal to
$(1-\gamma)\Phi^{-1}(1-1/n) + \gamma\Phi^{-1}(1-1/(e n))$, where
$\Phi^{-1}$ denotes the quantile function (inverse of the probability
distribution function) of the Gaussian distribution with mean 0 and
standard deviation 1, $e \approx \num{2.718282}$ is Euler's number,
and $\gamma \approx \num{0.5772157}$ is the Euler-Mascheroni
constant. This expected value increases with $n$, and it is already
1.8 for $n=16$.

Furthermore, when there are $n=16$ normalized residuals, and the data
are mutually consistent and the underlying statistical model applies,
the odds are better than even (53\,\% probability, in fact) that at
least one will have absolute value greater than 2. Therefore, and in
general, requiring that all normalized residuals, after application of
the expansion factor, should have absolute values less than 2 leads
to excessively large expansion factors.

\subsection{Additive models}
\label{sec:additive}

An alternative treatment, which indeed is the most prevalent approach
to blend independent measurement results, from medicine to metrology,
including when the results are mutually inconsistent, involves an
additive model for the measured values, of the form
\begin{equation}
  \label{eqn:additive}
  G_{j} = G + \lambda_{j} + \varepsilon_{j}.
\end{equation}

This model acknowledges the possibility that the different experiments
may be measuring different quantities, by introducing experiment
\emph{effects} $\{\lambda_{j}\}$ such that, given $\lambda_{j}$, the
expected value of $G_{j}$ is $G + \lambda_{j}$. The standard deviation
of the measurement error $\varepsilon_{j}$ is the reported
uncertainty, $u(G_{j})$. Since the experiment effects may be
indistinguishable from zero, this model can also accommodate mutually
consistent data.

On first impression, it may seem that the model is non-identifiable:
that by making $\lambda_{j}$ large and $\varepsilon_{j}$ small, or
vice-versa, the same value of $G_{j}$ may be reproduced. However, the
fact that the data are not only the $\{G_{j}\}$ but also the
$\{u(G_{j})\}$, resolves the potential ambiguity: since
the $\{\varepsilon_{j}\}$ are comparable to their corresponding
$\{u(G_{j})\}$, if the $\{G_{j}\}$ turn out to be appreciably more
dispersed than the $\{u(G_{j})\}$ intimate, then this suggests that
the $\{\lambda_{j}\}$ cannot all be zero.

The most common modeling assumption is that the $\{\lambda_{j}\}$ are
a sample from a Gaussian distribution with mean 0 and standard
deviation $\tau$, which quantifies the dark
uncertainty. \citet{koepke-2017a} discuss several variants of this
\emph{random effects} model, and describe procedures to fit them to
measurement data. Some of these procedures are implemented in the
\emph{NIST Consensus Builder}, which is a Web-based application
publicly and freely available at \url{https://consensus.nist.gov}
\citep{koepke-2017}.

The DerSimonian-Laird procedure to fit random effects models to
measurement data is used most commonly in meta-analysis in medicine
\citep{dersimonian-1986,dersimonian-2015}.  This procedure yields the
conventional weighted mean when the estimate of dark uncertainty is 0.

\begin{figure}
  \centering
  \includegraphics[keepaspectratio=true,width=\linewidth]%
  {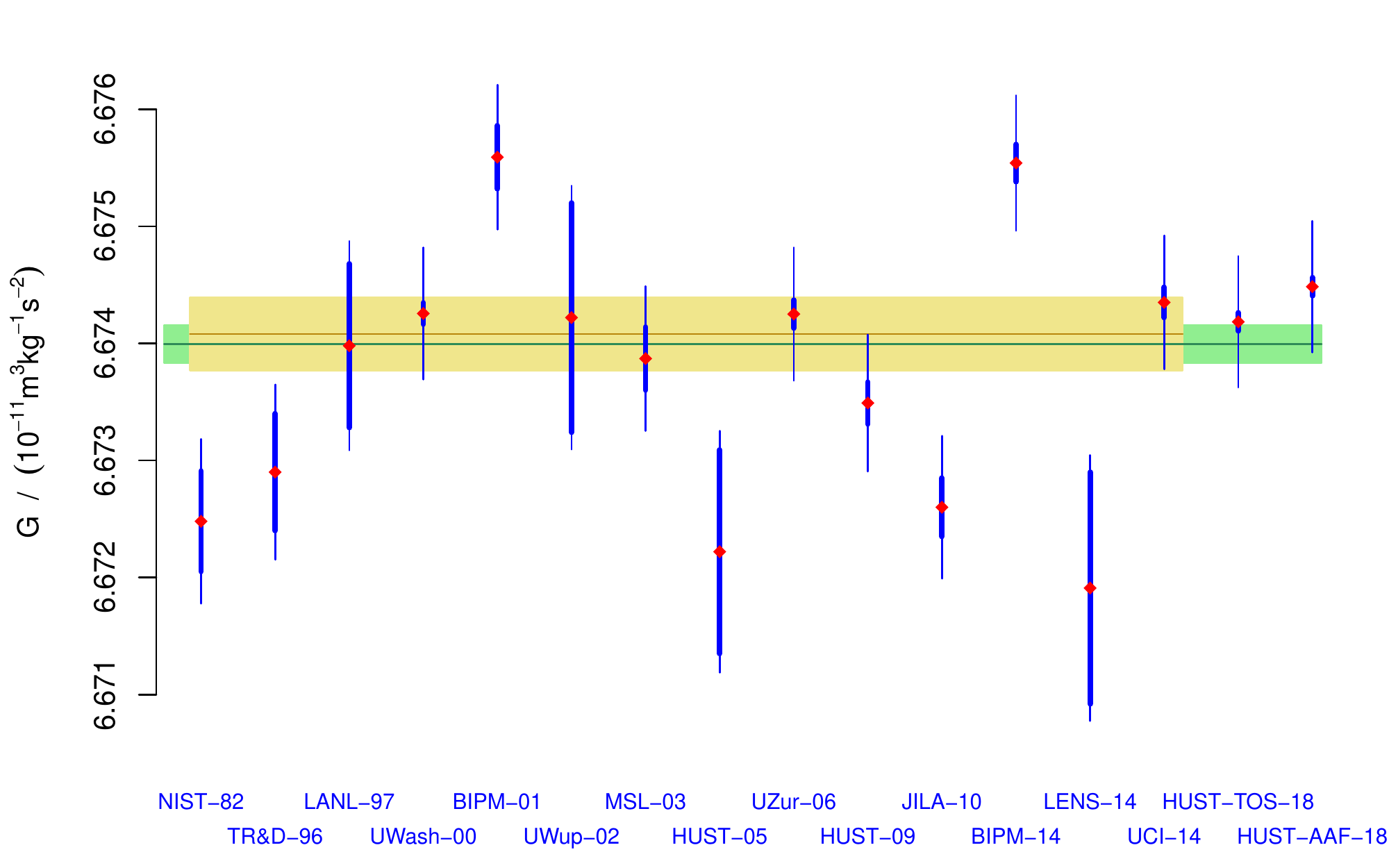}
  \caption{Measurement results from Table~\ref{tab:newtonian}, where
    the measured values are represented by red diamonds, and plus or
    minus one reported standard uncertainties are represented by thick
    vertical blue line segments centered at the measured values. The
    thin line segments that extend the thick segments indicate the
    contribution from dark uncertainty, corresponding to
    $G_{j} \pm (u^{2}(G_{j}) +
    \tau_{\text{DL}}^{2})^{\text{\textonehalf}}$. The horizontal green
    line represents the consensus value $G_{\text{DL}}$, and the light
    green band represents $G_{\text{DL}} \pm u(G_{\text{DL}})$. The
    horizontal brown line and light brown band are the counterparts of
    the green line and light green band, for the CODATA recommended
    value $G_{2014}$, which did not incorporate the measurement
    results from either \textsf{HUST-TOS-18} or \textsf{HUST-AAF-18},
    and used the uncorrected result from \textsf{JILA-10}.o Compare
    with Figure~\ref{fig:birge}.} \label{fig:dl}
\end{figure}

The version of the DerSimonian-Laird procedure implemented in the
\emph{NIST Consensus Builder} estimates $G$ as \SI{6.67399e-11}%
{\meter\tothe{3}\kilo\gram\tothe{-1}\second\tothe{-2}}, with
associated standard uncertainty $u(G) =$ \SI{0.00025e-11}%
{\meter\tothe{3}\kilo\gram\tothe{-1}\second\tothe{-2}} (including the
Knapp-Hartung adjustment \citep{knapp-2003}), and dark uncertainty
$\tau_{\text{DL}} = \SI{0.00056e-11}%
{\meter\tothe{3}\kilo\gram\tothe{-1}\second\tothe{-2}}$. These results
are depicted in Figure~\ref{fig:dl}, and appear in
Table~\ref{tab:comparison}, row DL.

\citet{rukhin-2011} propose a version of the random effects model
where both the experiment effects $\{\lambda_{j}\}$ and the
measurement errors $\{\varepsilon_{j}\}$ are samples from two
different Laplace distributions. The consensus value in this case is a
weighted median, \SI{6.67408e-11}%
{\meter\tothe{3}\kilo\gram\tothe{-1}\second\tothe{-2}}, with
associated standard uncertainty
\SI{0.00030e-11}{\meter\tothe{3}\kilo\gram\tothe{-1}\second\tothe{-2}}. The
corresponding estimate of dark uncertainty is
$\tau_{\text{LAP}} = \SI{0.00127e-11}%
{\meter\tothe{3}\kilo\gram\tothe{-1}\second\tothe{-2}}$
(Table~\ref{tab:comparison}, row LAP).

Several other versions of the additive random effects model are
implemented in various packages for the R environment for statistical
data analysis and graphics \citep{r-2018}, including: \texttt{metafor}
\citep{viechtbauer-2010} (used to produce the estimates of $G$ labeled
ML, MP, and REML in Table~\ref{tab:comparison}); \texttt{metaplus}
\citep{beath-2016} (for estimate STU in Table~\ref{tab:comparison});
and \texttt{metamisc} \citep{debray-2019} (for estimate MM in
Table~\ref{tab:comparison}), among many others.

\section{Shades of Dark Uncertainty}
\label{sec:shades}

A comparison of Figures~\ref{fig:birge} and \ref{fig:dl}, and of the
underlying models and corresponding numerical results, reveals
important and obvious differences, as well as two noteworthy
commonalities: (i) the consensus values, although numerically
different, neither differ significantly from one another once their
associated uncertainties are taken into account, nor do they differ
significantly from the 2014 CODATA recommended value, even though both
incorporate measurement results (\textsf{HUST-TOS-18} and
\textsf{HUST-AAF-18}) that were not yet available when this
recommended value was produced, as well as the corrected result for
JILA-10; (ii) both penalize the effective uncertainty of the
individual measurement results uniformly, albeit one differently from
the other.

The penalty applies regardless of how the measured values are situated
relative to the consensus value, and regardless also of whether the
reported uncertainties are small or large. For example, in
Figure~\ref{fig:dl} one might have expected \textsf{JILA-10} and
\textsf{BIPM-14} to have been penalized with appreciably larger
components of dark uncertainty than \textsf{UZur-06} or
\textsf{HUST-09}.

Figure~\ref{fig:birge} reveals other, possibly even less palatable
anomalies, which are specific to the multiplicative inflation of the
reported uncertainties: in particular, that the results from
\textsf{LANL-97}, \textsf{UWup-02}, \textsf{HUST-05}, and
\textsf{LENS-14}, end-up contributing essentially nothing to the
consensus value.

The pattern of the measurement results depicted in Figure~\ref{fig:dl}
is fairly typical: on the one hand, there is a cluster of results
(including \textsf{UZur-06} and \textsf{HUST-TOS-18}) that, all by
themselves, would be mutually consistent and indeed have measured
values that lie quite close to the consensus value; on the other hand,
there is another cluster (including \textsf{BIPM-01}, \textsf{JILA-10}
and \textsf{BIPM-14}) whose measured values lie much farther afield,
to either side of the consensus value.

To increase the flexibility of additive random effects models, in
particular to enable them to cope with such mixed bag of results, and
to alleviate the inequities arising from applying the same dark
uncertainty penalty to all the results, regardless of how they are
situated relative to the consensus value, we have developed a new
model that yields different evaluations of dark uncertainty for
different subsets of the measurement results. We call the
corresponding, different $\tau$s, \emph{shades of dark uncertainty}.

This new model, which we introduce in the next Section, represents the
probability distributions of the measured values as mixtures of
distributions, similarly to how the linear opinion pool, implemented
in the \emph{NIST Consensus Builder} \citep{koepke-2017a}, represents
them. (The results of applying the linear opinion pool to the data in
Table~\ref{tab:newtonian} are labeled LOP in
Table~\ref{tab:comparison}.)

For a simple example of a mixture, consider two dice: one is cubic
with faces numbered 1 through 6; the other is dodecahedral with faces
numbered 1 through 12; the faces of each die are equally likely to
land up when the die is rolled. Suppose that one die is chosen at
random so that the cubic die is twice as likely to be chosen as the
dodecahedral die, and then it is rolled. The probability distribution
of the outcome is a mixture of two discrete, uniform distributions:
the probability of a four is
$(2/3) \times (1/6) + (1/3) \times (1/12) = 5/36$.

And if one is told that a four turned up, but not which die was
rolled, then one can use Bayes rule \citep{degroot-2012,possolo-2011a}
to infer that it was the cubic die with
$((1/6) \times (2/3)) / (5/36) =$ 80\,\% probability. Given the
results of multiple realizations of this procedure (choosing a die at
random and rolling this die), one may then compute the probabilities
of the outcomes having originated in the cubic die. Those outcomes for
which this probability is greater than 50\,\% may be said to form one
cluster, and the others a different cluster.

\section{Bayesian mixture model}
\label{sec:mix}

The mixture model that we propose is parametric and Bayesian, and
depends on the number, $K$, of shades of dark uncertainty to be
entertained. ``Parametric'' means that all probability distributions
are determined by a finite number of scalar parameters. ``Bayesian''
means that the data ($\{(G_{j}, u(G_{j})\}$) are modeled as observed
values of random variables, that the unknowns (true value of $G$,
probabilities of membership in the latent clusters, and shades of dark
uncertainty) are modeled as non-observable random variables, and that
the information the data hold about the unknowns is extracted by
application of Bayes's rule and distilled into the \emph{posterior
  distribution} of the unknowns (which is the conditional distribution
of the parameters given the data).

Subsection~\ref{sec:model} characterizes the model given the number,
$K$, of components in the mixture, and Subsection~\ref{sec:selection}
describes how a value for $K$ is chosen automatically, from among the
models corresponding to $K=1, 2, \dots, n$, so that the procedure
produces the ``best'' model, according to a Bayesian model selection
criterion.

\subsection{Model definition}
\label{sec:model}

Mixture models do not actually partition the measured values into
clusters, each with its own shade of dark uncertainty. Instead,
each measured value belongs to all the latent clusters simultaneously,
but typically with rather different probabilities of belonging to each
one of them. This fuzzy reality notwithstanding, it is often a useful
simplification to say that a measured value belongs to the latent
cluster that it has the largest posterior probability of belonging
to. Accordingly, and to present the results vividly, in
Section~\ref{sec:results} we ``assign'' each measurement to the latent
cluster that the measurement has the largest posterior probability of
belonging to --- the so-called maximum \emph{a posteriori} estimate
(MAP) of cluster membership.

The $K$ distributions being mixed (which define the latent clusters)
are Gaussian, and they have different standard deviations, which are
the shades of dark uncertainty, $\tau_{1}, \dots, \tau_{K}$. The
results include an estimate of $G$, an evaluation of the associated
uncertainty, estimates of the $\{\tau_{k}\}$, as well as the
identification of the latent cluster that each measurement result most
likely belongs to.

Since the model is Bayesian and will be fit to the measurement results
via Markov Chain Monte Carlo (MCMC) \citep{gelman-2013}, not only
estimates and standard uncertainties, but also coverage (credible)
intervals, may easily be derived for all the parameters in the model:
$G$, the $\{\tau_{k}\}$, and the cluster membership probabilities
$\bm{\pi}_{j} = (\pi_{j,1}, \dots, \pi_{j,K})$, where
$\pi_{j,K} = 1-(\pi_{j,1}+\dots+\pi_{j,K-1})$, for $j = 1, \dots, n$,
and $\pi_{k,j}$ denotes the probability that measurement $j$ belongs
to cluster $k$, for $k=1,\dots, K$. Therefore, the model corresponding
to a particular value of $K$ has $1 + K + n(K-1)$ parameters.

The reported uncertainties $\{u(G_{j})\}$, even though they are data,
are treated as known quantities on the assumption that they are based
on infinitely many degrees of freedom. In cases where they are not,
the model can easily be modified to accommodate the finite numbers of
degrees of freedom that the $\{u(G_{j})\}$ may be based on.

The model is hierarchical \citep{gelman-2007}: (i) given $G$, the
$\{\tau_{k}\}$, and the $\{\bm{\pi}_{j}\}$, the measured values are
modeled as observed outcomes of Gaussian random variables, with
$G_{j}$ having a Gaussian distribution with mean $G$ and standard
deviation $\upsilon_{j}$ such that
\begin{equation}
  \label{eqn:upsilon}
  \upsilon^{2}_{j} = u^{2}(G_{j}) + \sum_{k=1}^{K} \pi_{j,k} \tau^{2}_{k},
\end{equation}
for $j=1,\dots,n$; (ii) $G$ has an essentially non-informative
Gaussian prior distribution with mean $G_{2014}$ and large variance;
(iii) the $\{\tau_{k}\}$ have mildly informative half-Cauchy
distributions whose medians have to be specified; and (iv) the
$\{\bm{\pi}_{j}\}$ have the same flat Dirichlet prior distribution
(all concentration parameters set equal to 1)
\citep[Chapter~49]{kotz-2000}. Furthermore, $G$, the $\{\tau_{k}\}$,
and the $\{\bm{\pi}_{j}\}$ are mutually independent \emph{a
  priori}. Equation~(\ref{eqn:upsilon}) makes precise the sense in
which the effective dark uncertainty for each measurement result is a
mixture of shades of dark uncertainty.

We implemented this model in the JAGS language \citep{plummer-2017},
and then used the implementation in R function \texttt{jags} defined
in package \texttt{R2jags} \citep{su-2015}, to produce samples from
the distribution of all the parameters via MCMC.

\subsection{Model selection}
\label{sec:selection}

Since the mixture representation of the dark uncertainty that appears
in the second term on the right-hand side of Equation~(1) involves
latent clusters and not a partition of the measurements into actual
clusters, in principle there is no constraint on the number, $K$, of
latent clusters. However, common sense dictates that there ought not
to be more than the number, $n$, of measurements being combined, hence
$1 \leqslant K \leqslant n$.

We consider the $n$ models corresponding to $K=1, \dots, n$ in turn,
and use each one to predict the value of $G$ that a future,
independent experiment may produce. The we choose the model that makes
the most accurate predictions. To be able to explain how this is done,
even if we omit all of the technical details, we need to introduce
some notation.

Let $D$ denote the data in hand ($n$ measured values and their
associated uncertainties), and $\theta$ denote the parameters in the
model defined in Subsection~\ref{sec:model}, with $K$ latent
clusters. Therefore, $\theta$ includes the unknown value of the
Newtonian constant of gravitation, $G$, the shades of dark uncertainty
$\{\tau_{k}\}$, and the probabilities, $\{\pi_{j,k}\}$, of membership
in the latent clusters. The probability density of the data given the
parameters is $f_{K}(D|\theta)$, and $p(\theta)$ is the prior
probability density of the parameters. The density of the posterior
distribution of the parameters given the data is given by Bayes's rule
\citep{degroot-2012}:
$q_{K}(\theta|D) = f_{K}(D|\theta) p(\theta) / g_{K}(D)$, where
$g_{K}(D) = \int f_{K}(D|\theta) p(\theta) \mathrm{d}\theta$, and the
integral is over the set of possible values of the parameters.

Our goal is to select the value of $K$ for which $h_{K}(D^{\ast}|D)$
is largest, where $D^{\ast}$ denotes a future measurement, and $h_{K}$
is the \emph{predictive posterior density} defined as
$h_{K}(D^{\ast}|D) = \int f_{K}(D^{\ast}|\theta) q_{K}(\theta|D)
\mathrm{d} \theta$ \citep{gelman-2013}. Since this future observation
$D^{\ast}$ is speculative (hence, unknown), the best we can do is
estimate $h_{K}(D^{\ast}|D)$ pretending that $D^{\ast}$ is one of the
results that we have, and that $D$ comprises all the results that we
have except that one.

For model selection, we rely on the Bayesian Leave-One-Out cross
validation score (LOO), which gauges the posterior predictive acumen
of the model under consideration. To compute it, the model is fitted
to $D_{-j}$ (all the measurements except the $j$th), and the
corresponding predictive density is evaluated at $D_{j}$ (the
measurement left out, here playing the role of future, independent
measurement), this process being repeated for $j=1,\dots,n$. Thus, for
each number of latent clusters $K$, the model is fitted $n$ times,
producing $n$ posterior densities $q_{-1,K}, \dots, q_{-n,K}$, each
based on $n-1$ measurements, and $\log h_{K}(D^{\ast}|D)$ is estimated
by the cross-validated \emph{predictive accuracy score}
\begin{equation}
  \label{eqn:score}
  \text{LOO} = \sum_{j=1}^{n}\log q_{-j,K}(D_{j}|D_{-j}),
\end{equation}
which we then transform into the LOO Information Criterion,
$\text{LOOIC} = -2 \times \text{LOO}$, which is numerically comparable
to Akaike's Information Criterion (AIC), a widely used
model selection criterion \citep{burnham-2004}.

Since determining each $q_{-j,K}$ involves an MCMC run, the procedure
outlined in the previous paragraph requires $nK$ MCMC runs. However, R
package \texttt{loo} \citep{vehtari-2019} offers a shortcut to this
onerous procedure and produces an approximation to the foregoing
average of values of log posterior densities using the results of a
single MCMC run.

Since the LOOIC involves the data and MCMC sampling, it is surrounded
by uncertainty, which we have evaluated using R function \texttt{loo}
defined in the package of the same name. In general, the smaller the
LOOIC, the better the model. However, differences between values of
LOOIC have to be interpreted taking their associated uncertainties
into account, as we will explain in Section~\ref{sec:results}.

\subsection{Similar models}
\label{sec:similar}

There is a growing collection of models whose purpose and devices are
similar to the model we described above. Here we mention only a few of
these alternatives.

\citet{burr-2005} describe a Bayesian semi-parametric model for
random-effects meta-analysis in the form of a Dirichlet mixture, which
is implemented in R package \texttt{bspmma} \citep{burr-2012}.

\citet{jara-2011} present Bayesian non-parametric and semi-parametric
models for a wide range of applications, including for linear,
mixed-effects models used in meta-analysis, using a Dirichlet process
prior distribution, or a mixture of Dirichlet process prior
distributions \citep{teh-2017}, for the distribution of the random
effects. Both R packages \texttt{DPpackage}
\citep{jara-2007,jara-2011} and \texttt{dirichletprocess}
\citep{ross-2018} facilitate the use of these priors.

\citet{jagan-2019} propose adjusting (typically inflating) each
reported uncertainty just enough to achieve mutual consistency, with
the adjustments obtained by minimization of a relative entropy
criterion. The results may be interpreted as involving estimates of
dark uncertainty that are tailored for each measurement result
individually.

Our proposal and \citet{rukhin-2019b}'s are similar in that
they both model the additional uncertainty directly, and not
through the distribution of the random effects as is done in most
other models. The main differences between our approach and
\citet{rukhin-2019b}'s are the following:
\begin{itemize}
\item Our mixture model comprises latent clusters, and each
  measurement may belong to all the clusters simultaneously, possibly
  with different probabilities, hence its effective dark uncertainty
  is a mixture of the shades of dark uncertainty of the latent
  clusters; \citet{rukhin-2019b} partitions the measurements into
  clusters and assigns a particular, same value of dark uncertainty to
  all the measurements in the same cluster.
\item \citet{rukhin-2019b} assumes that the measurements in one of the
  clusters are mutually consistent, hence that it has no dark
  uncertainty (the ``null'' cluster). In most cases there will be
  multiple clusters whose measurements are mutually consistent, and
  the results may depend on which one is chosen to play the role of
  ``null'' cluster.
\end{itemize}

\section{Results}
\label{sec:results}

Table~\ref{tab:selection} lists the values of the model selection
criterion LOOIC, and associated uncertainties, for the models
corresponding to $K=0,1,\dots,16$ shades of dark uncertainty. The case
with $K=0$ is the common mean model, $G_{j} = G + \varepsilon_{j}$
(cf. Equation~(\ref{eqn:additive})), which does not recognize dark
uncertainty, and is vastly inferior to the models that do recognize it.

As $K$ increases from 1 to $n$, the LOICC undergoes its largest drop
in value from $K=1$ to $K=2$, where it reaches its minimum, thus
suggesting that the best model should have $K=2$ latent
clusters. However, the large uncertainties associated with the LOOIC
caution that this choice is only nominally better than any other.

One of the reasons why the LOOIC does not achieve a sharp, deep
minimum, and instead keeps hovering near its minimum as $K$ increases
above 2, is that for some of the larger values of $K$, the number of
effective latent clusters is much smaller than $K$. For example, when
$K = 10$, there are only 5 different MAP estimates of cluster
``membership'', that is, 5 different effective latent clusters.  Next
we explain what we mean by ``effective latent clusters.''

In Subsection~\ref{sec:model} we pointed out that we ``assign'' each
measurement to the latent cluster that the measurement has the largest
posterior probability of belonging to --- the so-called maximum
\emph{a posteriori} estimate (MAP) of cluster membership: these MAP
assignments are reflected in the different colors of the labels in
Figures~\ref{fig:mix-pi} and \ref{fig:mix}.

Recognizing that the model with $K=2$, although nominally the best, is
not head and shoulders above the other models with $K \geqslant 1$, we
further invoke the general principle that, everything else being just
about comparable, one is well-advised to take the simpler model:
therefore, we will proceed on the assumption that the best model has
$K=2$ latent clusters. This choice is also supported by the fact that
the model with $K=2$ assigns clearly smaller amounts of dark
uncertainty to \textsf{UWash-00} and to \textsf{UZur-06} than to
results that are similarly precise, or even more precise, but lie
farther away from the consensus value. The model with $K=1$ would be
incapable of drawing such distinctions.

\begin{table}
  \centering 
    \begin{tabular}{lrrrrrrr}
      \toprule
      {$K$} & 0 & 1 & 2 & 3 & 4 & 5 & 6 \\
      {LOOIC} & $-26.7$ & $-170.8$ & $\mathbf{-173.9}$ & $-172.8$ 
                            & $-172.0$ & $-172.2$ & $-171.6$\\ 
      {$u(\text{LOOIC})$} & 74.0 & 5.2 & 8.0 & 7.7 & 7.9 & 7.9 & 8.0\\[0.65ex]
      \cmidrule{2-8}
      {$K$} & & 7 & 8 & 9 & 10 & 11 & 12 \\
      {LOOIC} & & $-171.2$ & $-171.6$ & $-171.2$ & $-171.2$ 
                                & $-171.2$ & $-171.1$\\
      {$u(\text{LOOIC})$} & & 8.2 & 8.1 & 8.2 & 8.3 & 8.3 & 8.3\\[0.65ex]
      \cmidrule{2-8}
      {$K$} & & 13 & 14 & 15 & 16 \\
      {LOOIC} & & $-170.9$ & $-170.6$ & $-170.7$ & $-170.6$\\
      {$u(\text{LOOIC})$} & & 8.5 & 8.5 & 8.5 & 8.5\\[0.65ex]
  \end{tabular}
  \caption{Values of the LOO Bayesian model selection criterion
    (LOOIC), for mixture models with $K=0,1,\dots,16$ latent clusters,
    fitted to the measurements listed in
    Table~\ref{tab:newtonian}. The column corresponding to $K=0$
    pertains to the common mean model, which does not recognize dark
    uncertainty. The best model has $K=2$ latent clusters, even if
    this suggestion is clouded by appreciable uncertainty,
    $u(\text{LOOIC})$.} \label{tab:selection}
\end{table}

The MCMC procedure yielded a sample of size \num{512000} drawn from
the joint posterior distribution of the parameters, resulting from
collating every 25th outcome from each of four chains of length
\num{4e6}, with burn-in of \num{8e5} iterations per chain. Each point
in this sample comprises one value for $G$, values for $\tau_{1}$ and
$\tau_{2}$, and cluster memberships $C_{1}, \dots, C_{n}$, and cluster
membership probabilities $\pi_{1,1}, \pi_{1,2}=1-\pi_{1,1}$, $\dots$,
$\pi_{n,1}, \pi_{n,2}=1-\pi_{n,1}$ for all the measurements in
Table~\ref{tab:newtonian}.

\begin{figure}
  \centering
  \includegraphics[keepaspectratio=true,width=0.75\linewidth]%
  {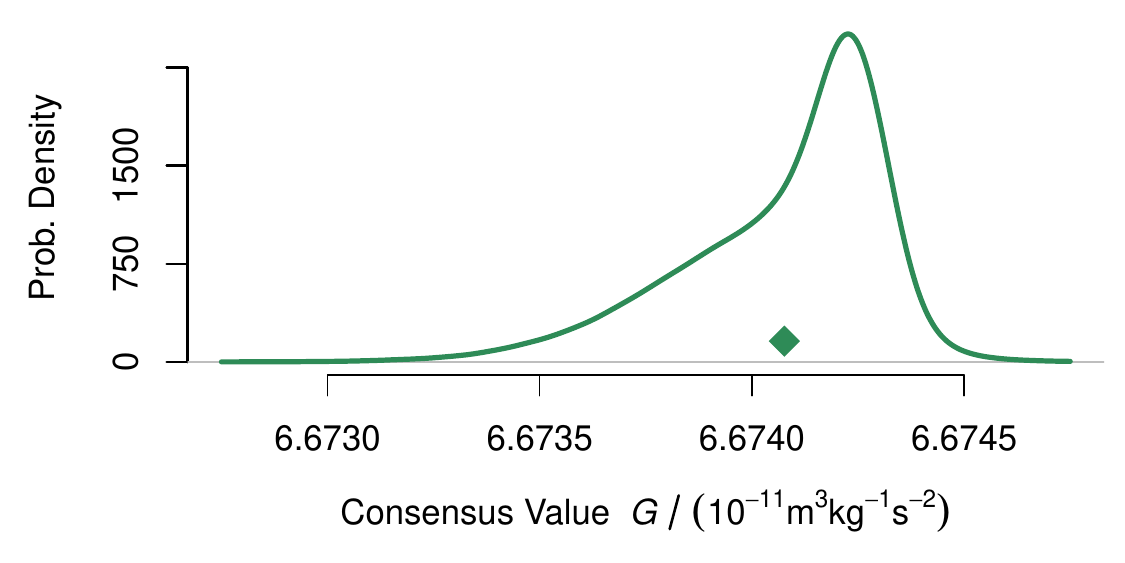} \\[1.5ex]
  \includegraphics[keepaspectratio=true,width=0.75\linewidth]%
  {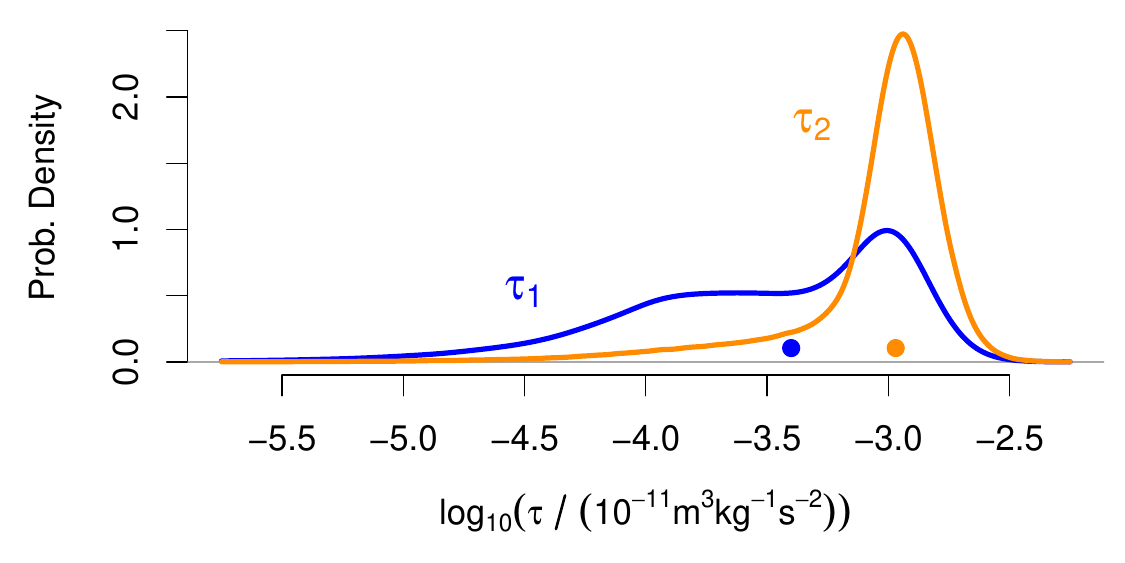}
  \caption{Density functions of the posterior probability
    distributions of $G$ (\textsc{upper panel}), and of the two shades
    of dark uncertainty, $\tau_{1}$ and $\tau_{2}$ (\textsc{lower
      panel}) for the model with $K=2$ latent clusters. The (green)
    diamond (\textsc{upper panel}) indicates the mean,
    \SI{6.67408e-11}{\meter\tothe{3}\kilo\gram\tothe{-1}\second\tothe{-2}},
    of the posterior distribution of $G$. The (blue and orange) dots
    (\textsc{lower panel}) indicate the medians of the posterior
    distributions of the two shades of dark uncertainty,
    $\widehat{\tau}_{1} =$
    \SI{0.0004}{\meter\tothe{3}\kilo\gram\tothe{-1}\second\tothe{-2}}
    and $\widehat{\tau}_{2} =$
    \SI{0.0011}{\meter\tothe{3}\kilo\gram\tothe{-1}\second\tothe{-2}}. Note
    the logarithmic scale of the horizontal axis in the lower
    panel.} \label{fig:mix-mutau}
\end{figure}

\begin{figure}
  \centering
  \includegraphics[keepaspectratio=true,width=0.85\linewidth]%
  {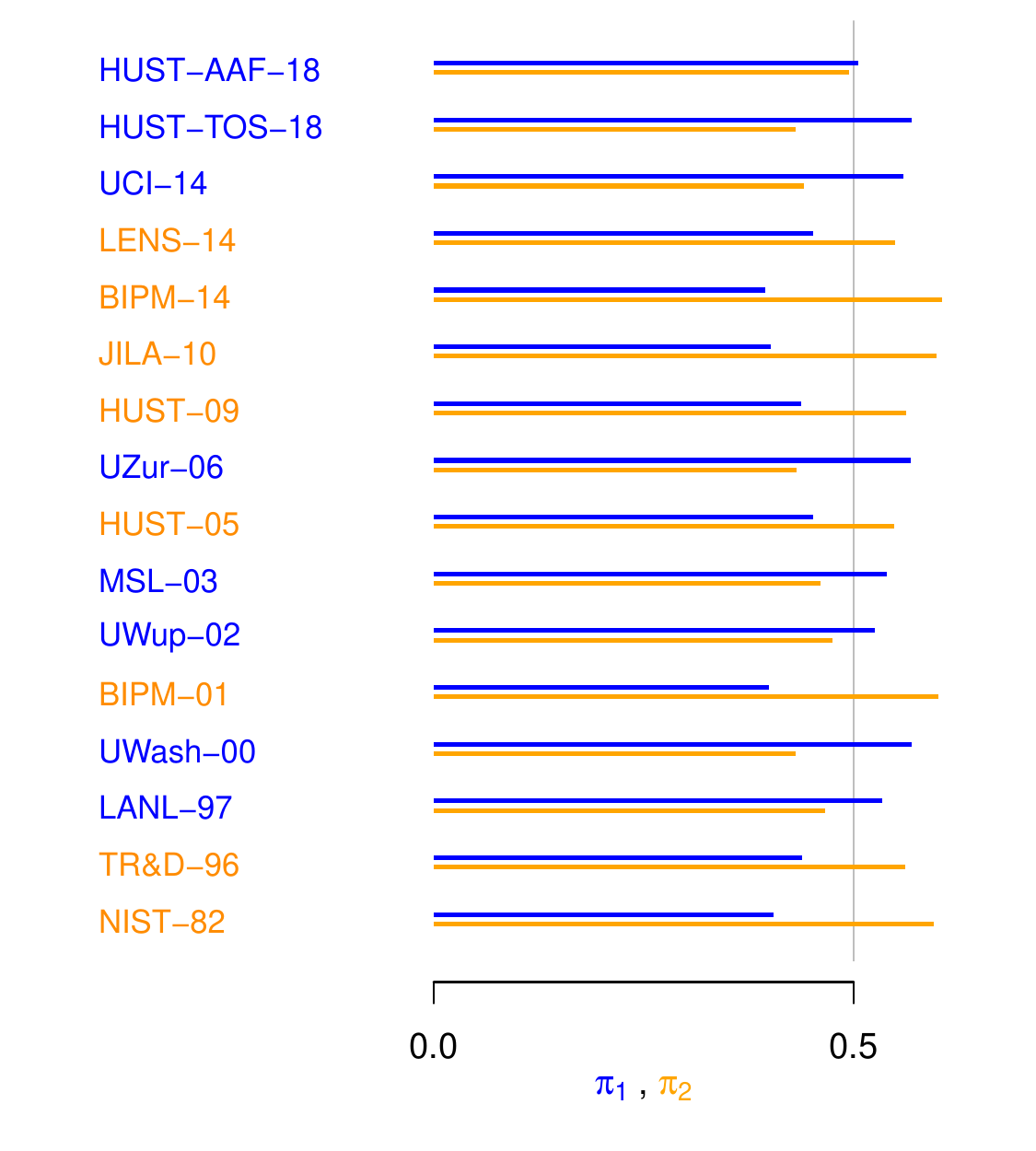}
  \caption{The two horizontal bars (the upper one, blue, for cluster
    1; the lower one, orange, for cluster 2) adjacent to each label
    represent the median posterior probabilities of membership in the
    two clusters. The horizontal bar that crosses the vertical, gray
    line at 0.5, determines the MAP estimate of cluster membership,
    denoted by the color of the label. The colors correspond to those
    used in the lower panel of
    Figure~\ref{fig:mix-mutau}.} \label{fig:mix-pi}
\end{figure}

\begin{figure}
  \centering
  \includegraphics[keepaspectratio=true,width=\linewidth]%
  {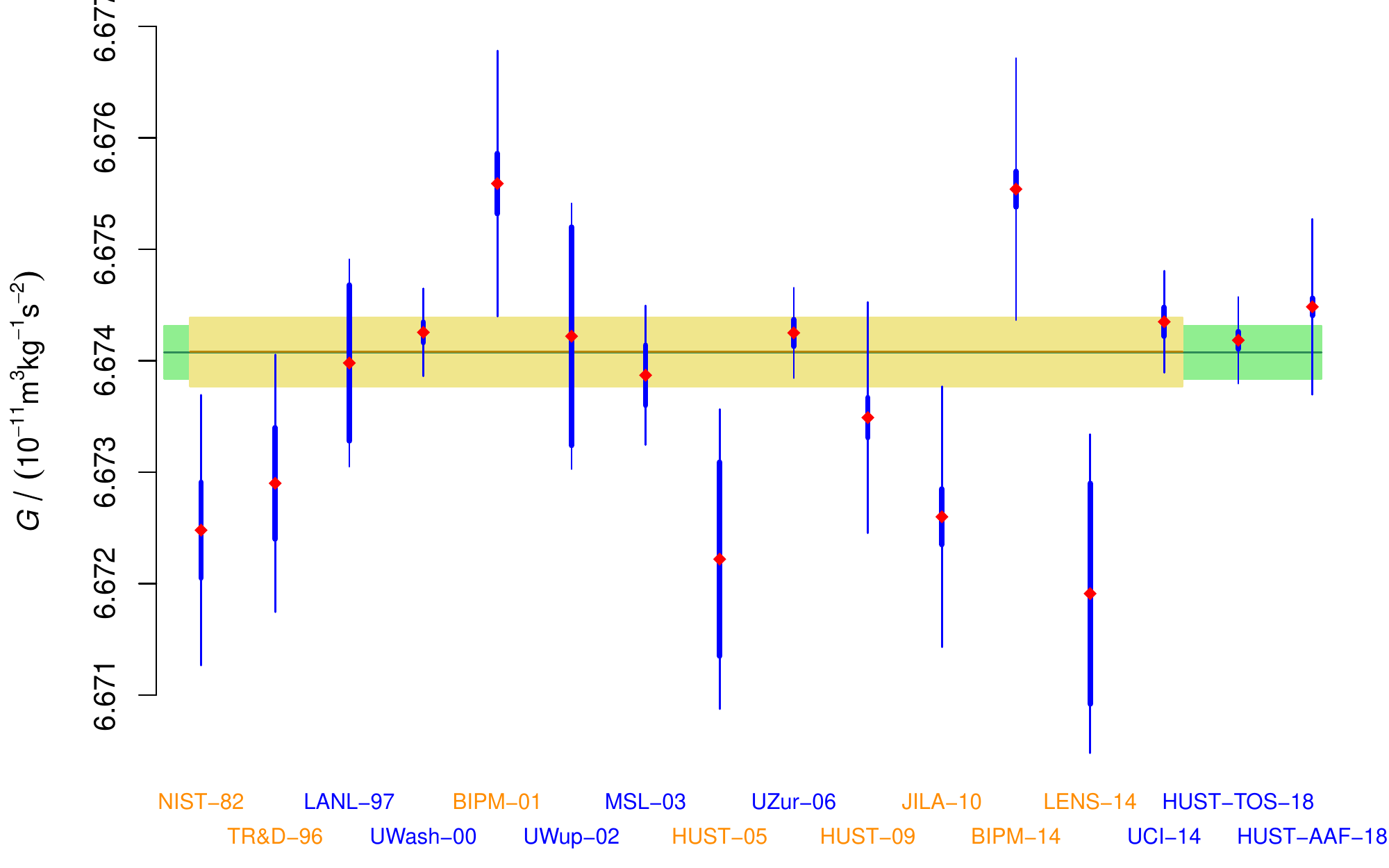}
  \caption{Results of fitting the Bayesian mixture model with two
    clusters, to the measurement results from
    Table~\ref{tab:newtonian}. The measured values are represented by
    red diamonds, and the measured values plus or minus the reported
    standard uncertainties are represented by thick vertical blue line
    segments centered at the measured values. The thin line segments
    that extend the thick segments indicate the contribution from the
    mixture of the two shades of dark uncertainty in the model,
    possibly of different size for the different measurements, but
    generally larger for those whose labels are highlighted in
    orange. The horizontal green line represents the consensus value,
    and the light green band represents the associated standard
    uncertainty. The horizontal brown line and light brown band are as
    in Figure~\ref{fig:dl}. The colors of the labels at the bottom
    indicate the more likely clusters that the different measurement
    results belong to: these colors are the same that are used in
    Figures~\ref{fig:mix-mutau} and \ref{fig:mix-pi}. Since cluster
    membership is determined by the distance from the measured value
    to the consensus value gauged in terms of the reported standard
    uncertainty, the dark uncertainty assigned to \textsf{HUST-AAF-18}
    is more comparable to the dark uncertainty assigned to
    \textsf{NIST-82} than to the dark uncertainty assigned to
    \textsf{HUST-TOS-18}.} \label{fig:mix}
\end{figure}

The upper panel of Figure~\ref{fig:mix-mutau} depicts the posterior
distribution of $G$. The Bayesian estimate of the consensus value was
chosen as the mean of the sample drawn from the posterior distribution
of $G$,
\SI{6.67408e-11}{\meter\tothe{3}\kilo\gram\tothe{-1}\second\tothe{-2}},
and the associated standard uncertainty,
\SI{0.00024e-11}{\meter\tothe{3}\kilo\gram\tothe{-1}\second\tothe{-2}},
as the standard deviation of the same sample. The 2.5th and 97.5th
percentiles of this sample are the endpoints of a 95\,\% coverage
(credible) interval for the true value of $G$ --- their values are
listed in the row of Table~\ref{tab:comparison} labeled BMM.

The lower panel of Figure~\ref{fig:mix-mutau} depicts the posterior
distributions of the two shades of dark uncertainty, $\tau_{1}$ and
$\tau_{2}$. Their Bayesian estimates,
$\widehat{\tau}_{1} =
\SI{0.0004}{\meter\tothe{3}\kilo\gram\tothe{-1}\second\tothe{-2}}$ and
$\widehat{\tau}_{2} =
\SI{0.0011}{\meter\tothe{3}\kilo\gram\tothe{-1}\second\tothe{-2}}$,
were chosen as the medians of their respective MCMC samples because
their distributions are markedly asymmetrical (lower panel of
Figure~\ref{fig:mix-mutau}), with very long right tails.

Figure~\ref{fig:mix-pi} depicts the medians of the posterior
probabilities of cluster membership, showing that for only a few of
the measurement results (for example, \textsf{BIPM-14} and
\textsf{UZur-06}) is membership in one of clusters clearly more likely
than membership in the other. \textsf{HUST-AAF-18} is just about as
likely to belong to one cluster as to the other, the difference
favoring membership in cluster 1 (which has the smallest shade of dark
uncertainty) by the narrowest of margins.

This fact helps explain why, as shown in Figure~\ref{fig:mix}, the
dark uncertainty assigned to \textsf{HUST-AAF-18} is closer to the
dark uncertainty assigned to \textsf{NIST-82} than to the dark
uncertainty assigned to \textsf{HUST-TOS-18}, even though, on the one
hand, the standard uncertainty reported for \textsf{HUST-AAF-18} is
quite similar to the standard uncertainty reported for
\textsf{HUST-TOS-18}, and on the other hand \textsf{HUST-AAF-18} lies
much closer to the consensus value than \textsf{NIST-82}. The reason
is that cluster membership is determined by the distance to the
consensus value gauged in terms of the reported standard uncertainty:
from this viewpoint \textsf{HUST-AAF-18} is just about as far from the
consensus value as \textsf{NIST-82}, and so much farther from it than
\textsf{HUST-TOS-18}.

Figure~\ref{fig:mix} depicts the data and the results of fitting
our mixture model to them. The meaning of the thick and thin vertical
blue lines is similar to the meaning that they have in
Figure~\ref{fig:dl}. A word of explanation is in order for how the
lengths of the thin lines were determined. The thin line centered at
$G_{j}$ represents $G_{j} \pm \upsilon_{j}$, where $\upsilon_{j}$ was
defined in Equation~(\ref{eqn:upsilon}). However, this is not how
the $\{\upsilon_{j}\}$ were computed.

The approach we took for computing $\upsilon_{j}$ tracks the actual
way in which MCMC unfolds, as closely as possible: each time the MCMC
process generates an acceptable sample, it provides a cluster
membership $k_{j} \in \{1,2\}$ for $G_{j}$, and produces also values
for $\tau_{1}$ and $\tau_{2}$ for the latent clusters: the
corresponding value of $\upsilon^{2}_{j}$ is
$u^{2}(G_{j}) + \tau^{2}_{k_{j}}$. The value used for $\upsilon_{j}$
in this Figure is the square root of the median of the \num{512000}
samples of $\upsilon^{2}_{j}$ computed as just described, for each
$j=1,\dots,n$.

\section{Conclusions}
\label{sec:conclusions}

The mixture model introduced in Section~\ref{sec:mix}, driven by the
model selection criteria outlined in Subsection~\ref{sec:selection},
by and large achieved one of the central goals of this contribution: to impart
flexibility to the conventional laboratory effects model, in
particular addressing successfully the long-standing grievance
resulting from penalizing (with dark uncertainty) measured values
lying close to the consensus value as severely as those that lie
farther afield.

This model truly excels in producing shades of dark uncertainty that
are finely attuned to the structure of the data, in particular to how
the measured values are arranged relative to one another and relative
to the consensus value, while taking their associated uncertainties
into account, as Figure~\ref{fig:mix} shows. Furthermore, the model
does all this without widening the uncertainty associated with the
consensus value.

Table~\ref{tab:comparison} summarizes consensus values for $G$, and
expressions of the associated uncertainty, that were derived from the
set of measurement results listed in Table~\ref{tab:newtonian}, by
various methods described in the foregoing, and in particular by the
new method that we have described (denoted BMM in this table).

\begin{table}
  \centering
  {\fontsize{10}{12}\selectfont
  \begin{tabular}{lllllcll} \toprule
    & $G/\gamma$ & $u(G)/\gamma$ & $\text{Lwr95}/\gamma$ 
    & $\text{Upr95}/\gamma$ && $\tau/\gamma$ & \\ \midrule\midrule
    \multicolumn{6}{l}{\textsc{multiplicative model --- birge's
    approach}} & \\ \midrule
    BRE  & 6.67429 & 0.00014 &         &         && & \\ 
    BRM  & 6.67429 & 0.00013 & 6.67403 & 6.67455 && & \\
    BRQ  & 6.67429 & 0.00011 &         &         && & \\ 
    MTE  & 6.67429 & 0.00015 &         &         && & \\
    \midrule\midrule          
    \multicolumn{6}{l}{\textsc{additive model --- conventional}} & \\ \midrule
    DL   & 6.67399 & 0.00025 & 6.67346 & 6.67453 && 0.00056& \\        
    ML   & 6.67390 & 0.00025 & 6.67341 & 6.67439 && 0.00091& \\        
    MP   & 6.67380 & 0.00060 & 6.67263 & 6.67497 && 0.00235& \\        
    REML & 6.67389 & 0.00026 & 6.67339 & 6.67440 && 0.00095& \\        
    STU  & 6.67390 &         & 6.67335 & 6.67440 && 0.00091& \\        
    \midrule\midrule                                                   
    \multicolumn{6}{l}{\textsc{additive model --- bayesian}} & \\ \midrule
    BG   & 6.67389 & 0.00027 & 6.67333 & 6.67442 && 0.00095& \\        
    LAP  & 6.67408 & 0.00030 & 6.67345 & 6.67471 && 0.00127& \\        
    MM   & 6.67389 & 0.00029 & 6.67327 & 6.67443 && 0.00101& \\        
    \midrule\midrule                                                   
    \multicolumn{6}{l}{\textsc{mixture model}}                         
    & $\tau_{1}/\gamma$ & $\tau_{2}/\gamma$ \\ \midrule               
    BMM  & 6.67408 & 0.00024 & 6.67350 & 6.67440 && 0.0004 & 0.0011\\
    LOP  & 6.67377 & 0.00117 & 6.67127 & 6.67577 && & \\
    \bottomrule
  \end{tabular}}
\caption{Consensus values, standard uncertainties, and 95\,\% coverage
  intervals (Lwr95, Upr95) for $G$, and estimates of shades of dark
  uncertainty ($\tau$, or $\tau_{1}$ and $\tau_{2}$ for BMM) produced
  by different statistical models and methods of data reduction. BRE
  $=$ Birge's approach with inflation factor that makes Cochran's $Q$
  equal to its expected value. BRM $=$ Birge's approach with inflation
  factor equal to its maximum likelihood estimate. BRQ $=$ Birge's
  approach with smallest inflation factor that makes data consistent
  according to Cochran's $Q$ test. MTE $=$ Weighted average after
  expansion of standard uncertainties to achieve normalized residuals
  with absolute value less than 2. DL $=$ DerSimonian-Laird with
  Knapp-Hartung adjustment. ML $=$ Gaussian maximum likelihood. MP $=$
  Mandel-Paule. REML $=$ Restricted Gaussian maximum likelihood. STU
  $=$ Random effects are a sample from a Student's $t$
  distribution. BG $=$ Bayesian hierarchical model from the \emph{NIST
    Consensus Builder} \citep{koepke-2017}, with estimate of $\tau$
  set to the median of its posterior distribution. LAP $=$ Laboratory
  effects and measurement errors modeled as samples from Laplace
  distributions \citep{rukhin-2011}. MM $=$ Bayesian model using a
  non-informative Gaussian prior distribution for $G$ and a uniform
  prior distribution for the dark uncertainty, implemented in R
  function \texttt{uvmeta} defined in package \texttt{metamisc}
  \citep{debray-2019}. LOP $=$ Linear opinion pool from the \emph{NIST
    Consensus Builder} \citep{koepke-2017}.} \label{tab:comparison}
\end{table}

It should be noted that the four variants of the multiplicative model
(BRE, BRM, BRQ, and MTE), all produce slightly larger estimates of $G$
than the additive models and the mixture models. Both LAP
\citep{rukhin-2011} and MP \citep{mandel-1970,paule-1982} yield
evaluations of dark uncertainty appreciably larger than the other
methods. The Bayesian mixture model (BMM) produces modest shades of
dark uncertainty because it capitalizes on smart, ``soft'' clustering
of the measurement results to explain the overall dispersion of the
measured values above and beyond what their associated, reported
uncertainties suggest.

In this contribution we have argued in favor of model-based approaches
to consensus building (as opposed to \emph{ad hoc} approaches like
those that are driven by the Birge ratio or by the sizes of the
absolute values of normalized residuals), particularly when faced with
measurement results that are mutually inconsistent.

Additive laboratory random effects models using mixtures, like the
model we introduced in Section~\ref{sec:mix}, seem especially promising as they
are able to identify subsets of results that appear to express
different shades of dark uncertainty, and then weigh them differently
in the process of consensus building, yet without disregarding the
information provided by any of the results being combined.

We have assembled in Table~\ref{tab:comparison} the results produced
by an assortment of methods not only to provide perspective on the new
method we are proposing (BMM), but also because arguably any of these
methods could reasonably be selected by different professional
statisticians working in collaboration with physicists engaged in the
measurement of $G$. Even though the underlying models and specific
validating assumptions differ, choosing one or another reflects mostly
inessential differences in training, preference, and experience in
statistical data modeling and analysis.

However, the variability of these estimates of $G$, which is
attributable solely to differences in approach, model selection, and
data reduction technique, amounts to about 78\,\% of the median of the
values of $u(G)$ listed in the same table, and to about 20\,\% of the
median of the values of $\tau$.

In other words, the statistical ``noise'' (that is, the vagaries and
incidentals that would lead one researcher to opt for a particular
model and method of data reduction, and another for a different model
and method) is clearly not negligible. Therefore, the development,
dissemination, and widespread adoption of best practices in
statistical modeling and data analysis for consensus building will be
contributing factors in reducing the uncertainty associated with any
consensus value that may be derived from an ever growing set of
reliable measurement results for $G$ obtained by increasingly varied
measurement methods.

Finally, we express our belief that one feature already apparent in
the collection of measurement results assembled in
Table~\ref{tab:newtonian} provides the greatest hope yet for appreciable
progress in the years to come: the fact that fundamentally different,
truly independent measurement methods have been employed, relying on
different physical laws, and yet there has been convergence toward a
believable consensus, even if it  still falls short of achieving a
reduction in relative uncertainty to levels comparable to what
prevails for other fundamental constants.

\section*{Acknowledgments}

The authors are much indebted to their NIST colleagues Eite Tiesinga
(Quantum Measurement Division, Physical Measurement Laboratory, and
Joint Quantum Institute, University of Maryland), and Amanda Koepke
(Statistical Engineering Division, Information Technology Laboratory),
who read an early draft, uncovered errors and pointed out confusing
passages, and provided many valuable comments and suggestions for
improvement. The authors are very grateful to Hartmut Petzold for
allowing them to share results of his historical research prior to
publication in a forthcoming book.



\end{document}